\documentclass[aps,prx,a4paper,superscriptaddress,twocolumn,amsmath,amssymb,longbibliography]{revtex4-1}

\usepackage{bm}
\usepackage{graphicx,epsfig}
\usepackage[usenames]{color}
\usepackage[english]{babel}

\usepackage{array}
\usepackage{dsfont}
\usepackage{multirow}
\usepackage{csquotes}
\usepackage{enumitem}

\usepackage{orcidlink}

\MakeOuterQuote{"}
\allowdisplaybreaks

\begin{document}

\newcommand{\E}{\mathcal{E}}
\newcommand{\G}{\mathcal{G}}
\newcommand{\Lag}{\mathcal{L}}
\newcommand{\M}{\mathcal{M}}
\newcommand{\N}{\mathcal{N}}
\newcommand{\U}{\mathcal{U}}
\newcommand{\R}{\mathcal{R}}
\newcommand{\F}{\mathcal{F}}
\newcommand{\V}{\mathcal{V}}
\newcommand{\C}{\mathcal{C}}
\newcommand{\I}{\mathcal{I}}
\newcommand{\s}{\sigma}
\newcommand{\up}{\uparrow}
\newcommand{\dw}{\downarrow}
\newcommand{\h}{\hat{H}}
\newcommand{\himp}{\hat{h}}
\newcommand{\g}{\mathcal{G}^{-1}_0}
\newcommand{\D}{\mathcal{D}}
\newcommand{\A}{\mathcal{A}}
\newcommand{\projs}{\hat{\mathcal{S}}_d}
\newcommand{\proj}{\hat{\mathcal{P}}_d}
\newcommand{\K}{\textbf{k}}
\newcommand{\Q}{\textbf{q}}
\newcommand{\T}{\tau_{\ast}}
\newcommand{\io}{i\omega_n}
\newcommand{\eps}{\varepsilon}
\newcommand{\+}{\dag}
\newcommand{\su}{\uparrow}
\newcommand{\giu}{\downarrow}
\newcommand{\0}[1]{\textbf{#1}}

\newcommand{\og}[1]{\textcolor{blue}{[\textbf{OG}: #1]}}
\newcommand{\sg}[1]{\textcolor{orange}{[\textbf{SG}: #1]}}
\newcommand{\dagga}{{\phantom{\dagger}}}
\newcommand{\ca}{c^{\phantom{\dagger}}}
\newcommand{\cc}{c^\dagger}

\newcommand{\pa}{{p}^{\phantom{\dagger}}}
\newcommand{\pc}{{p}^\dagger}

\newcommand{\fa}{f^{\phantom{\dagger}}}
\newcommand{\fc}{f^\dagger}

\newcommand{\da}{{d}^{\phantom{\dagger}}}
\newcommand{\dc}{{d}^\dagger}

\newcommand{\bF}{\mathbf{F}}
\newcommand{\bD}{\mathbf{D}}

\newcommand{\bR}{\mathbf{R}}
\newcommand{\bQ}{\mathbf{Q}}
\newcommand{\bq}{\mathbf{q}}
\newcommand{\bqp}{\mathbf{q'}}
\newcommand{\bk}{\mathbf{k}}
\newcommand{\bh}{\mathbf{h}}
\newcommand{\bkp}{\mathbf{k'}}
\newcommand{\bp}{\mathbf{p}}
\newcommand{\bL}{\mathbf{L}}
\newcommand{\bRp}{\mathbf{R'}}
\newcommand{\bx}{\mathbf{x}}
\newcommand{\bX}{\mathbf{X}}
\newcommand{\by}{\mathbf{y}}
\newcommand{\bz}{\mathbf{z}}
\newcommand{\br}{\mathbf{r}}
\newcommand{\Ima}{{\Im m}}
\newcommand{\Rea}{{\Re e}}
\newcommand{\Pj}[2]{|#1\rangle\langle #2|}
\newcommand{\ket}[1]{\vert#1\rangle}
\newcommand{\bra}[1]{\langle#1\vert}
\newcommand{\setof}[1]{\left\{#1\right\}}
\newcommand{\fract}[2]{\frac{\displaystyle #1}{\displaystyle #2}}
\newcommand{\Av}[2]{\langle #1|\,#2\,|#1\rangle}
\newcommand{\av}[1]{\langle #1 \rangle}
\newcommand{\Mel}[3]{\langle #1|#2\,|#3\rangle}
\newcommand{\Avs}[1]{\langle \,#1\,\rangle_0}
\newcommand{\eqn}[1]{(\ref{#1})}
\newcommand{\Tr}{\mathrm{Tr}}

\newcommand{\bba}{b^{\phantom{\dagger}}}
\newcommand{\bbc}{b^\dagger}

% \title{Accelerated Simulations of Strongly Correlated Matter by Machine-Learning the Ghost Gutzwiller Embedding Variational Space}

\title{Linear Foundation Model for Quantum Embedding: Data-Driven Compression of the Ghost Gutzwiller Variational Space}

\author{Samuele Giuli}
\affiliation{Center for Computational Quantum Physics, Flatiron Institute, New York, New York 10010, USA}

\author{Hasanat Hasan}
\affiliation{School of Physics and Astronomy, Rochester Institute of Technology, Rochester, New York 14623, USA}

\author{Benedikt Kloss}
\affiliation{Center for Computational Quantum Physics, Flatiron Institute, New York, New York 10010, USA}

\author{Marius S. Frank}
\affiliation{Department of Chemistry, Aarhus University, 8000, Aarhus C, Denmark}

\author{Tsung-Han Lee}
\affiliation{Department of Physics, National Chung Cheng University, Chiayi 62102, Taiwan}

\author{Olivier Gingras}
\affiliation{Center for Computational Quantum Physics, Flatiron Institute, New York, New York 10010, USA}
\affiliation{Université Paris-Saclay, CNRS, CEA, Institut de physique théorique, 91191, Gif-sur-Yvette, France}

\author{Yong-Xin Yao}
\altaffiliation{Corresponding author: ykent@iastate.edu}
\affiliation{Ames National Laboratory, Ames, Iowa 50011, USA}
\affiliation{Department of Physics and Astronomy, Iowa State University, Ames, Iowa 50011, USA}

\author{Nicola Lanat\`a} %\orcid{0000-0003-0003-4908}}
\altaffiliation{Corresponding author: nxlsps@rit.edu}
\affiliation{School of Physics and Astronomy, Rochester Institute of Technology, Rochester, New York 14623, USA}
\affiliation{Center for Computational Quantum Physics, Flatiron Institute, New York, New York 10010, USA}

\date{\today}

\begin{abstract} 

Simulations of quantum matter rely mainly on Kohn--Sham density functional theory (DFT), which often fails for strongly correlated systems. Quantum embedding (QE) theories address this limitation by mapping the system onto an auxiliary embedding Hamiltonian (EH) describing fragment--environment interactions, but the EH is typically large and its iterative solution is the primary computational bottleneck.
We introduce a linear foundation model for QE that utilizes principal component analysis (PCA) to compress the space of quantum states needed to solve the EH within a small variational subspace. Using a data-driven active-learning scheme, we learn this subspace from EH ground states and reduce each embedding solve to a deterministic ground-state eigenvalue problem in the reduced space.
Within the ghost Gutzwiller approximation (ghost-GA), we show for a three-orbital Hubbard model that a variational space learned on a Bethe lattice is transferable to square and cubic lattices without additional training, while substantially reducing the cost of the EH step. We further validate the approach on plutonium, where a single variational space reproduces the energetics of all six crystalline phases while reducing the cost of the EH solution by orders of magnitude. This provides a practical route to overcome the main computational bottleneck of QE frameworks, paving the way for high-throughput \textit{ab initio} simulations of strongly correlated materials at a near-DFT cost.

\end{abstract}

\maketitle

\section{Introduction}

Theoretical and computational tools capable of simulating solids and molecules ``ab-initio''---based solely on the quantum-mechanical interactions between their elementary constituents---are deeply integrated across scientific disciplines, from condensed matter physics and materials science to chemistry and biochemistry. By connecting emergent behaviors to microscopic origins, these tools enable interdisciplinary research and the development of new materials~\cite{NSTC2011,Chen2015}. 
Kohn-Sham density functional theory (DFT) serves as the standard workhorse for many such calculations~\cite{HohenbergandKohn,KohnandSham}.
However, strong electron correlations~\cite{Kotliar-Science} can drastically influence the electronic and structural properties of materials, particularly in systems containing $3d$ transition metals, lanthanides, and actinides~\cite{Eschrig1990, corr-structure-1,corr-structure-2,corr-structure-3,corr-structure-4,corr-structure-5,corr-structure-6, npj-lanata,Cross2019,Galley2018}. 
These strong electron correlations, related to the ``multi-configurational problem'' in chemistry, underlie critical phenomena including magnetism and high-temperature superconductivity~\cite{highTc-1,highTc-2,highTc-3,highTc-4,highTc-5,highTc-6,highTc-7,highTc-8}.

Quantum embedding (QE) theories provide a framework to calculate the electronic structure of extended strongly correlated systems. 
In QE methods, the interacting lattice or molecular problem is mapped onto a local fragment (impurity) coupled to an environment described by an embedding Hamiltonian (EH), which encodes how the fragment exchanges particles and correlations with the rest of the system. 
Examples include DFT combined with dynamical mean-field theory (DMFT)~\cite{dmft_book,Anisimov_DMFT,Held-review-DMFT,DMFT,xidai_impl_LDA+DMFT,LDA+U+DMFT}, density matrix embedding theory (DMET)~\cite{DMET,DMET-spectral}, rotationally invariant slave-boson theory (RISB)~\cite{Georges,rotationally-invariant_SB,Lanata2016}, the multi-orbital Gutzwiller approximation (GA)~\cite{Gutzwiller3,Our-PRX,lanata-barone-fabrizio,GA-infinite-dim,Fang,Ho,Gmethod,Bunemann,Attaccalite}, and the ghost-Gutzwiller approximation (ghost-GA)~\cite{Ghost-GA,ALM_g-GA,gRISB,gDMET}.
These methods have demonstrated predictive power for correlated materials, but their widespread application in fields such as materials science and computational catalysis is hindered by a severe computational bottleneck:
the solution of the EH. In iterative workflows such as charge self-consistency loops or structural optimizations, the EH parameters are updated recursively, which requires solving a quantum many-body impurity problem at each step. This makes QE calculations orders of magnitude more demanding than standard DFT.

Machine learning (ML) has emerged as a promising route to eliminate this bottleneck. In the context of ML-assisted DMFT, significant progress has been made in learning the mapping involving the infinite-dimensional hybridization function and the dynamical Green's function~\cite{surrogate-Millis,Surrogate-sheridan,surrogate-Sturm, surrogate_Ren_2021,Venturella2023ACS,Agapov2024arxiv,Deng2025arxiv,Lee2025PRB} or the impurity self-energy~\cite{Dong2024PRB,Mitra2025chemrxiv}. These approaches address the difficult problem of learning a mapping between continuous functions of frequency, which formally constitute infinite-dimensional input and output spaces.
Within the class of QE methods such as DMET, RISB, and the ghost-GA, by contrast, the physical system is mapped to an EH coupled to a \textit{finite} number of bath sites, and the output consists of static single-particle expectation values, specifically the density matrix. This implies that the solution to the embedding problem is entirely encoded in the ground-state energy function of the EH, as the required physical outputs can be obtained directly as gradients of this energy.
Leveraging this structure, previous works have attempted to learn the ground-state energy of the EH directly~\cite{Our-ML-actinides,surrogate-Marius}. However, such direct learning approaches do not automatically enforce fundamental physical constraints, such as the requirement that density matrix eigenvalues lie between zero and one (Pauli principle). Consequently, achieving high accuracy and stability requires extensive training data and complex sampling strategies to ensure the learned landscape remains physical.

In this work, we present a framework that overcomes these limitations by shifting the learning target from the EH energy function to the low-dimensional structure of its ground states. We hypothesize that the physically relevant ground states of the EH reside on a low-dimensional manifold embedded within the exponentially large Hilbert space, and we show that this manifold can be efficiently learned from data using the principal component analysis (PCA).
This procedure enables us to determine a linear variational space, which constitutes a "foundation model" for QE frameworks, offering two decisive advantages. First, by projecting the EH onto a learned subspace, we compress the EH many-body problem, reducing it to the diagonalization of a small matrix and eliminating the computational bottleneck. Second, because the method retains the structure of a variational ansatz, physical constraints such as the Pauli principle are satisfied by construction.

We validate this approach within the ghost-GA framework, by benchmarking it against the three-orbital Hubbard model, demonstrating remarkable transferability where a solver trained on a Bethe lattice accurately solves square and cubic lattices without retraining. Furthermore, we apply the method to the complex phase diagram of plutonium (Pu), reproducing the energetics of all six crystalline phases using a single variational space. 
Finally, we propose a pathway for extending our data-driven framework to larger EHs, using ground-state data generated by solvers such as matrix product states (MPS)~\cite{itensor,block2,DMRG-REVIEW,DMRG-original-White-PRL,DMRG-original-White-PRB,DMRG_PhysRevB.104.115119}, neural quantum states (NQS), variational impurity solvers based on superpositions of Gaussian states~\cite{Bauer-impurity-solver,Wu2025_gaussian}, or quantum-assisted methods~\cite{Sriluckshmy2025,AVQITE,Error-mitigation-GPR,Chen2025}.

\section{The ghost-GA Algorithmic Structure}
\label{subsec:gGA}

The ghost-GA theoretical framework is rooted in the variational principle and, similarly to DMFT, in the idea of the infinite-dimension limit~\cite{GA-infinite-dim}. 
Specifically, it is based on a variational wavefunction represented as $|\Psi_G\rangle = \hat{P}_G|\Psi_0\rangle$, where $\hat{P}_G$ is an operator acting on a reference single-particle wavefunction (Slater determinant) $|\Psi_0\rangle$. 
The wavefunction $|\Psi_0\rangle$ is defined in an auxiliary Hilbert space, where the number of fermionic modes is larger than in the physical Hilbert space by a tunable factor $B$.
This factor $B$ governs the size of the variational space and the accuracy of the method. The operator $\hat{P}_G$ maps states from the auxiliary space into the physical space, modifying the weights of the local electronic configurations of all fragments. Both $|\Psi_0\rangle$ and $\hat{P}_G$ are optimized to minimize the variational energy, evaluated neglecting all contributions that vanish in the limit of infinite-dimensional systems (as in DMFT). Therefore, like DMFT, the ghost-GA can accurately capture dynamical correlations, but not spatial correlations.

In the context of real-material calculations, the union of DFT with a many-body method is required to capture strong electronic correlations.
The algorithmic structure of DFT+ghost-GA, resulting from a convenient reformulation of the energy minimization problem for the multi-configurational wavefunction $|\Psi_G\rangle$~\cite{Our-PRX,Ghost-GA,ALM_g-GA}, closely mirrors that of DFT+DMFT: it features a charge self-consistent iterative procedure nested within the solution of an EH---which serves as the impurity model.
Physically, the EH describes how each atomic fragment interacts with its environment, ensuring that the feedback between local electronic correlations and the global electron density is consistently accounted for throughout the simulation.

It is important to note that this algorithmic structure applies to both real materials (in DFT+ghost-GA) and theoretical models (such as the Hubbard model).
In fact, like DFT+DMFT, the DFT+ghost-GA theory is ultimately encoded in a modified DFT functional, transforming the standard one-body Kohn-Sham (KS) reference system into a multi-orbital Hubbard model with local interactions, termed the ``KS-Hubbard Hamiltonian'' (KSH). Similar to conventional DFT, the core of DFT+QE is a charge self-consistency loop, which continuously updates the KSH system. The KSH model is then solved with the QE method of choice. Thus, real-material calculations require a recursive QE solution of the EH, with an additional charge self-consistency loop nested within the process.

The EH of each fragment $i$ is defined in second quantization as follows (omitting the label $i$ for simplicity):
\begin{align}
\hat{H}^{\text{EH}} &= \hat{H}^{\text{int}} 
+ \sum_{\alpha,\beta=1}^{\nu} E_{\alpha\beta} \, c^\dagger_{\alpha}c_{\beta}
\label{eq:gendefEH}
\\
&+ \sum_{a=1}^{B\nu} \sum_{\alpha=1}^{\nu} \left(D_{a\alpha}\, c^\dagger_{\alpha}b_{a} + \text{H.c.}\right) 
+ \sum_{a,b=1}^{B\nu} \Lambda_{ab} \,b^\dagger_{a}b_{b}
\nonumber
\,,
\end{align}
where $c^\dagger_\alpha $ ($c_\alpha $) is the creation (annihilation) operator for an embedded electron with spin-orbit index $\alpha$, $b^\dagger_a$ ($b_a$) is the creation (annihilation) operator for an electron with index $a$ in the bath and $\nu$ denotes the number of spin-orbitals in the fragment (2 for $s$, 6 for $p$, 10 for $d$, and 14 for $f$-shells).
The first term, $\hat{H}^{\text{int}}$, captures the local electronic interactions within the fragment:
%\vspace{-0.5cm}
\begin{equation}
    \hat{H}^{\text{int}} = 
    \frac{1}{2}\sum_{p,q,r,s=1}^{2l+1}\sum_{\sigma,\sigma'\in{\uparrow,\downarrow}} \mathcal{V}_{pqrs} \,
    c^\dagger_{p\sigma} c^\dagger_{r\sigma'} c_{s\sigma'} c_{q\sigma}
    \,,
    \label{Hemb-general}
\end{equation}
where $l=0,1,2,3$ corresponds to $s$, $p$, $d$, and $f$ electronic shell, and $\mathcal{V}_{pqrs}$ are 
the interaction coefficients.
The second term corresponds to the one-body part of the fragment (impurity), with $E$ specifying its one-body energies.
The subsequent terms in the Hamiltonian describe the coupling of the fragment to an auxiliary ``bath'' of $B\nu$ orbitals, governed by hybridization parameters, $D$, and bath energies, $\Lambda$.
The parameter $B$ controls the balance between computational complexity and accuracy: $B=1$ recovers the standard multi-orbital GA~\cite{Gutzwiller3,Our-PRX,lanata-barone-fabrizio,GA-infinite-dim,Fang,Ho,Gmethod,Bunemann,Attaccalite}, while $B=3$ has been systematically shown to achieve DMFT-level accuracy for ground-state properties~\cite{Ghost-GA, ALM_g-GA,TH1,TH2,TH3}, at a substantially lower computational cost.

From now on, we assume that the Coulomb integrals $\mathcal{V}_{pqrs}$ are fully encoded in the Hubbard-repulsion parameters $U$ and the Hund's coupling constant $J$~\cite{Anisimov_1997}, setting the Hubbard $U$ as energy unit.
For later convenience, we introduce the following compact notation:
\begin{align}
\hat{H}^{\text{EH}}(\mathbf{X}) &= 
\hat{H}_0+
\sum_{r=1}^{\dim(\mathbf{X})}X_r\hat{H}_r
\,,
\end{align}
where $\hat{H}_0$ is the term corresponding to $U=1$, while $J/U$ and the entries of $E/U$, $D/U$ and $\Lambda/U$ are stored in a real vector $\mathbf{X}$, which is updated self-consistently throughout the ghost-GA loop.
We denote $|\Phi(\mathbf{X})\rangle$ the ground state of $\hat{H}^{\text{EH}}(\mathbf{X})$.

The output parameters to be computed at each iteration of the ghost-GA algorithmic structure are the following ground-state single-particle density matrix elements: 
\begin{align}
    {h}_{\alpha a}(\mathbf{X}) &= \langle\Phi(\mathbf{X})|c_{\alpha}^\dagger b_{a}|\Phi(\mathbf{X})\rangle \,,
    \label{eq:V_1bdm} 
    \\
    {d}_{mn}(\mathbf{X}) &= \langle\Phi(\mathbf{X})|b_{m}^\dagger b_{n}
    |\Phi(\mathbf{X})\rangle
    \,. 
    \label{eq:delta_1bdm}
\end{align}

\subsection{Computational complexity of ghost-GA}

The ghost-GA algorithmic structure outlined above has two key computational advantages:
\begin{enumerate}
\item In ghost-GA, it is sufficient to solve an EH with a relatively small bath ($B = 3$) to obtain the total energy of the lattice model with an accuracy comparable with DMFT, which in principle requires a bath with infinite degrees of freedom.
\item While in DMFT, the output $\mathbf{Y}$ is the local Green's function of the fragment $i$, the ghost-GA requires only the iterative calculation of the entries of the ground-state single-particle density matrix.
\end{enumerate}
Because of these reasons, each ghost-GA iteration can be tackled with a wealth of efficient state-of-the-art methods that do not suffer from the so-called continuous-time quantum Monte Carlo fermionic ``sign problem''~\cite{ctqmc-sign-problem_PhysRevB.101.045108,ctqmc-sign-problem_PhysRevLett.126.216401}. 
Typical choices are the following:
\begin{enumerate}[label=(\roman*)]
    \item For $s$ and $p$ orbitals, the $B=3$ EH can be solved using exact diagonalization methods (Arnoldi iterations). 
    \item For $d$-orbital systems, the EH can be solved employing MPS techniques~\cite{itensor,block2,DMRG-REVIEW,DMRG-original-White-PRL,DMRG-original-White-PRB,DMRG_PhysRevB.104.115119}.
    \item Recently, quantum computing methods have also shown promise~\cite{Sriluckshmy2025,AVQITE,Error-mitigation-GPR,Chen2025}. 
\end{enumerate}

The accuracy and algorithmic structure of the ghost-GA framework has proven useful within a broad range of challenging strongly-correlated problems~\cite{Altermagnetism-gGA,Guerci,gGA-PhysRevB.107.235150,gGA-PhysRevB.111.125110,D4FD00053F,Pasqua2025},
and provides us with the opportunity to tackle simulations otherwise impossible. 
This notably includes simulations of low-symmetry systems where the self-energy has a non-diagonal structure, such as multi-orbital systems with disorder or coexistence of crystal/ligand-field splittings and spin-orbit coupling. Another exciting example is computational catalysis, where strongly correlated materials have been shown to induce new mechanisms, potentially paving the way for more efficient ammonia production under mild conditions~\cite{Ammonia-Norskov, zhang2024spin}.
Meanwhile, even the ghost-GA would be too computationally expensive for extensive high-throughput calculations.

\subsection{Challenge of designing an EH solver taking full advantage of the ghost-GA algorithmic structure}

Despite the favorable algorithmic advantages discussed above, the computational cost of solving the EH remains the primary bottleneck within the self-consistency ghost-GA cycle, limiting the scale and throughput of simulations with respect to mean-field approaches such as Kohn-Sham DFT or DFT+U and, therefore, their applicability for large-scale materials discovery.
This motivates us to address the following key question:
{Can this computational bottleneck be eliminated by developing a more efficient solver directly leveraging the ghost-GA algorithmic structure, thereby achieving DMFT-like accuracy at a near-DFT cost?}

In this work, we introduce a pathway to achieve this goal by leveraging on the dimensionality reduction for learning an optimized variational space from data, systematically reducing the computational overhead of EH computations by orders of magnitude.

\begin{figure}
  \centering
  \includegraphics[width=0.48\textwidth]{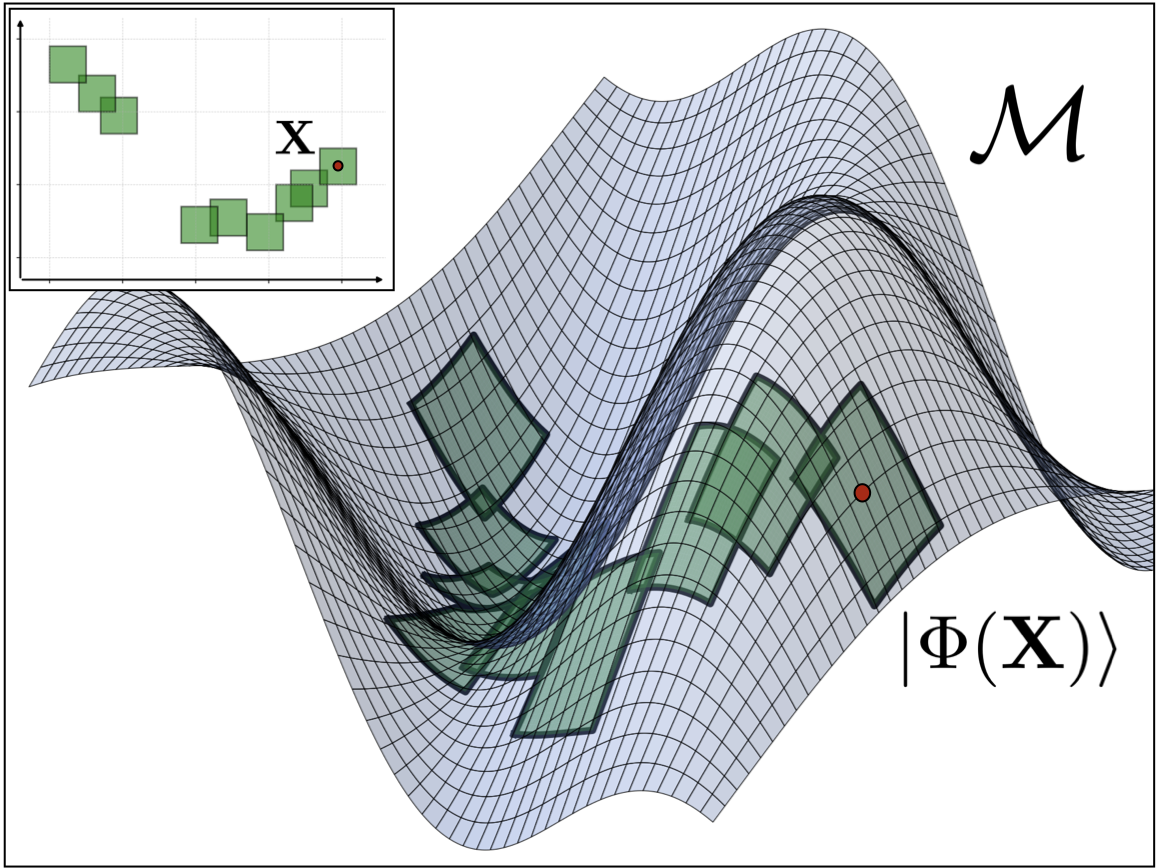}
  \caption{Geometric representation of the ground-state manifold $\mathcal{M}$. The curved surface represents the manifold of ground states $|\Phi(\mathbf{X})\rangle$. The manifold is partitioned into local regions (green rectangles), each corresponding to a domain in the Hamiltonian parameter space $\mathbf{X}$, which is schematically illustrated in the inset.}
  \label{figure1}
\end{figure}

\section{Overview of previous applications of ML in QE methods}
\label{subsec:ML}

The foundation of this work lies in a key observation made in Refs~\cite{surrogate-Millis,Our-ML-actinides, Surrogate-sheridan,surrogate-Sturm,surrogate-Marius,surrogate_Ren_2021}: within all QE methods, the form of the EH is universal. That is, its mathematical structure depends solely on the impurity's structure (typically representing $d$-electron or $f$-electron shells) and remains independent of the specific material or molecule containing it. 
Based on this observation, the goal is to train a machine to learn once and for all the universal map between the embedding parameters $\mathbf{X}$ and the corresponding outputs $\mathbf{Y}$, for each impurity type. Note that, for a given bath size $B$, this requires to train only \textit{one} machine for each type of impurity (e.g., one for $d$ electrons and one for $f$ electrons).
This approach aims to bypass the computationally expensive quantum-mechanical solution of the EH, allowing DFT+QE simulations of any solid and molecule at a DFT-like cost.

The perspective of bypassing the computational bottleneck of QE simulations of strongly correlated methods motivated extensive work along the lines depicted above, particularly focused on DMFT. In that case, $\mathbf{X}$ includes the hybridization function of a fragment, its on-site energies, the Hubbard parameter $U$, and Hund's constant $J$, while $\mathbf{Y}$ represents the corresponding fragment's Green's function. This input-output structure demands learning from a function of infinite variables (the hybridization function), which involves aspects such as bath parametrization and fitting. Similarly, it involves the problem of learning a continuous output (the local Green's function of the impurity). So far, research in this field has focused on identifying optimal parametrizations for these infinite-dimensional inputs and outputs. However, to date, benchmarks have focused only on the single-band impurity models with a \textit{fixed} mathematical form, such as a featureless (flat) hybridization function. The scalability of ML-based impurity solvers to full DMFT self-consistent calculations---where the hybridization function is iteratively updated, and does not retain any predetermined mathematical form---is yet to be proven, even for single-orbital lattice models.

As explained above in Sec.~\ref{subsec:gGA}, the ghost-GA introduces a notably simpler input-output structure with respect to DMFT. Here, the EH has a finite bath, defining the parameters with a finite set, and the output parameters are encoded in the gradient of the EH ground-state energy function:
\begin{align}
\mathcal{E}(\mathbf{X}) &= \langle\Phi(\mathbf{X})|\hat{H}^{\text{EH}}(\mathbf{X})|\Phi(\mathbf{X})\rangle
\label{eq:universal-energy}
\\
\mathbf{Y}(\mathbf{X}) &= \nabla\mathcal{E}(\mathbf{X})
\,,
\end{align}
where $\Phi(\mathbf{X})$ is the ground state of $\hat{H}^{\text{EH}}(\mathbf{X})$.

For example, the partial derivative of the EH ground-state energy with respect to $[\Lambda]_{aa}$ is 
$\langle\Phi(\mathbf{X})| b_{a}^\dagger b_{a}|\Phi(\mathbf{X})\rangle$. Therefore, if we could learn once and for all the universal energy function $\mathcal{E}(\mathbf{X})$, along with its gradient for all types of fragments $i$ (particularly $s$, $p$, $d$, or $f$ electronic shells, corresponding to $\nu_i = 2,6,10,14$ of Eq.~\eqref{Hemb-general}, respectively), we would be able to perform ghost-GA calculations of any material and molecule, bypassing altogether the computational bottleneck of calculating the ground-state of the EH. This mathematical structure motivated previous attempts to directly learn the function $\mathcal{E}(\mathbf{X})$ using ML frameworks~\cite{surrogate-Marius}. However, these approaches do not automatically enforce all qualitative physical constraints that the energy landscape must satisfy. 
One such example is that the density matrix elements in Eqs~\eqref{eq:V_1bdm} and \eqref{eq:delta_1bdm}, encoded in the gradient of the energy, must satisfy $N$-representability constraints (e.g., the Pauli principle)~\cite{N-Representability-Coleman}. 
The absence of this physical information in the model required generating extensive training data and complex sampling strategies to achieve sufficient accuracy~\cite{surrogate-Marius}, limiting scalability to complex materials.

Here we propose a different approach that resolves these problems. Instead of directly learning the energy landscape, we introduce a method to construct a linear variational space for the EH ground state directly from data. This strategy effectively compresses the relevant Hilbert space, similarly to how image compression identifies and retains only the essential features of a picture while discarding redundant information. By solving the problem within this optimized subspace, the framework enforces $N$-representability constraints by construction, yielding a drastically more compact, accurate, and scalable approach.

\section{Learning an Efficient Variational Space for the EH}
\label{Sec:learning}

The set of all relevant ground states $|\Phi(\mathbf{X})\rangle$ forms a $\dim(\mathbf{X})$-dimensional manifold $\mathcal{M}$ within the full Hilbert space $\mathcal{H}$, as represented in Fig.~\ref{figure1}. 
Note that this is a direct consequence of the ghost-GA's finite-bath structure. 
The fact that $\mathcal{M}$ lives in such a low-dimensional corner of the Hilbert space presents a unique opportunity: if we could learn an efficient variational space capturing the states in $\mathcal{M}$, we could construct a new type of EH solver.
%%%%%%

The key difficulty to address is how to explicitly construct this latent space, given that $\mathcal{M}$ is not known explicitly.
This makes our goal distinct from general-purpose variational frameworks where the structure of the ansatz is fixed a priori by physical considerations (e.g., MPS).

Our solution is based on the idea of learning the variational space directly from data, a concept also explored in other quantum many-body physics methods aimed at dimension reduction, such as eigenvector continuation~\cite{Eigenvector-Continuation_PhysRevLett.121.032501} and foundation models based on neural networks~\cite{Rende-foundation-NNQS,Zhou2025NQS-impurity,Ma-DMET-NNQS}.
Specifically, we realize a data-driven program constructing a piecewise-linear variational space, as follows:
\begin{itemize}
    \item We partition the space of $\mathbf{X}$ parameters into local domains, each representing a chart of the ground-state manifold $\mathcal{M}$, see Fig.~\ref{figure1}.
    
    \item For each local domain, we construct from data a linear variational space with tunable dimension $K$ (controlling the resulting accuracy).
\end{itemize}

We structure our explanation as follows:
In Sec.~\ref{subsec:varspaceEH} we show that, once a suitable low-dimensional linear subspace has been specified within a given region of $\mathcal{M}$, the EH problem can be reformulated and solved within that reduced space.
In Sec.~\ref{subsec:construction-var-space} we then show how this variational subspace can be constructed from data by PCA compression of EH ground states, and we place this construction on a principled footing by interpreting it as a linear foundation model for the family of EHs.

\subsection{Variational space construction of the EH}
\label{subsec:varspaceEH}

Let $\mathcal{V}_K$ be a $K$-dimensional linear variational space for a given region of the EH ground state manifold $\mathcal{M}$, and let $\{u_1,\ldots,u_K\}$ be an orthonormal basis of $\mathcal{V}_K$.
We define $\Pi_K$ as the orthogonal projector onto $\mathcal{V}_K$.
The problem of extremizing the corresponding variational energy is equivalent to calculating the ground state of the following $K \times K$ matrix:
\begin{equation}
[\tilde{H}^K(\mathbf{X})]_{nm} = [\tilde{H}_0^K]_{nm} + \sum_{r=1}^{\dim(\mathbf{X})}X_r[\tilde{H}_r^K]_{nm}
\,,
\label{eq:projHk1}
\end{equation}
where the matrix elements 
\begin{equation}
[\tilde{H}_r^K]_{nm} = \langle u_n|\hat{H}_r|u_m\rangle
\label{eq:compressedHK}
\end{equation}
can be conveniently pre-calculated and stored for all $r=0,\ldots,\dim(\mathbf{X})$.
In particular, the resulting variational approximation to the universal energy function previously introduced in Eq.~\eqref{eq:universal-energy} is the following:
\begin{align}
\mathcal{E}(\mathbf{X}) \approx \mathcal{E}_K(\mathbf{X}) & = \sum_{n,m=1}^{K}c_n^*(\mathbf{X})\langle u_n|\hat{H}^{\text{EH}}(\mathbf{X})|u_m\rangle c_m(\mathbf{X})
\nonumber\\
&= \lambda_{\min}\!\left(\tilde{H}_0^K + \sum_{r=1}X_r\tilde{H}_r^K\right)
\,,
\label{eq:projHk2}
\end{align}
where $\lambda_{\min}$ denotes the lowest eigenvalue of $\tilde{H}^K(\mathbf{X})$, and $c_m(\mathbf{X})$ are the components of the corresponding eigenvector:
\begin{equation}
    \ket{\Phi_K(\mathbf{X})}=\sum_{k=1}^K c_k(\mathbf{X})\,\ket{u_k}
    \,.
\end{equation}
The gradient of $\mathcal{E}_K(\mathbf{X})$ is available through the Hellmann-Feynman theorem. For example:
\begin{align}
\frac{\partial\mathcal{E}_K(\mathbf{X})}{\partial\Lambda_{aa}} &= \sum_{n,m=1}^{K}c_n^*(\mathbf{X})\langle u_n| b_{a}^\dagger b_{a}|u_m\rangle c_m(\mathbf{X}) 
\,.
\end{align}

This framework has three fundamental properties: 
\begin{enumerate}
    \item Since this representation is based on the variational principle, the resulting density matrices satisfy all $N$-representability conditions by construction~\cite{N-Representability-Coleman}. For example, the occupation numbers are guaranteed to lie within physical bounds,
    $0\leq \langle {\Phi_K(\mathbf{X})}| 
    b_{a}^\dagger b_{a}
    |{\Phi_K(\mathbf{X})}\rangle\leq 1$, 
    strictly enforcing the Pauli principle.
    
    \item Partitioning the parameter space in multiple charts reduces the complexity of the learning problem for each individual chart, and the cost of evaluating observables decreases as a function of the variational-space truncation $K$.
    
    \item The partitioning strategy also provides us with the option of focusing only on training our model within physically-realizable regions of the parameters space, which, as we are going to show, is considerably more efficient and scalable.
\end{enumerate}

It is important to point out that the possibility of calculating an efficient linear variational approximation to the EH problem is theoretically possible only thanks to the finite size of the EH bath, as for an infinite bath Anderson's orthogonality catastrophe~\cite{Anderson_Infrared-Catastrophe} would render ground states for different parameters mutually orthogonal, precluding the possibility of approximating them within any finite $K$.

\subsection{Construction of variational space for each chart of $\mathcal{M}$}
\label{subsec:construction-var-space}

Let us consider any given chart of $\mathcal{M}$, represented in Fig.~\ref{figure1}.
We aim to identify the basis vectors of $\{u_1,\ldots,u_K\}$ spanning the corresponding "optimal" variational space $\mathcal{V}_K$, of any dimension $K$ smaller than the Hilbert space dimension $M_B$.

This optimization problem can be formalized mathematically as follows.
We denote $\Pi_K$ as the orthogonal projector over $\mathcal{V}_K$. Consider a probability distribution $p(\mathbf{X})$ over the parameter space for the Hamiltonians $\hat{H}(\mathbf{X})$. This induces a corresponding distribution $P(\Phi)$ over their respective ground states $\Phi(\mathbf{X})$. We aim to calculate, for any given integer $K$, the orthogonal projector $\Pi_K$ that minimizes the following quantity:
\begin{equation}
C(\Pi_K) = E_P\left[\|(\Pi_K-I)\Phi\|^2\right]
\,,
\end{equation}
where $I$ is the identity and $E_P$ denotes the expectation value with respect to $P(\Phi)$. 
In practice, we approximate $C(\Pi_K)$ using a finite number $N$ of samples $\{\mathbf{X}^1,\ldots,\mathbf{X}^N\}$, assuming a uniform (constant) probability distribution $p(\mathbf{X})$:
\begin{equation}
C(\Pi_K) \simeq \frac{1}{N}\sum_{s=1}^{N}\|(\Pi_K-I)\Phi(\mathbf{X}^s)\|^2
\,,
\end{equation}
where we are introducing a superscript $s$ to indicate the $s$-th sampled vectors $\mathbf{X}^s$, with components $[\mathbf{X}^s]_r$.

To solve this optimization problem, we first construct the data matrix containing the sampled ground states as columns:
$\mathbb{M} = \left[\Phi(\mathbf{X}^1) | \Phi(\mathbf{X}^2) | \dots | \Phi(\mathbf{X}^N)\right]$.
We then perform a singular value decomposition (SVD) on this matrix:
\begin{equation}
\mathbb{M} = U \Sigma V^\dagger = \sum_{a=1}^{N} \sigma_a |u_a\rangle \langle v_a|\,, \label{eq:M_SVD}
\end{equation}
where \( U \) and \( V \) are unitary matrices containing the left and right singular vectors, respectively, and \( \Sigma \) is the diagonal matrix of singular values \( \sigma_a \), sorted in descending order.
A standard result from PCA states that the leading singular vectors provide the optimal low-rank approximation to the data. Applying this to our case, we obtain that the optimization problem of minimizing \( C(\Pi_K) \) is solved by the orthogonal projector:
\begin{equation}
\Pi_K = \sum_{a=1}^{K} |u_a\rangle \langle u_a|\,, \label{eq:PIK_projector}
\end{equation}
which minimizes the projection error in approximating the sampled ground states. 
In conclusion, our variational space is $\mathcal{V}_K = \text{Span}(\{u_1, \dots, u_K\})$ and the resulting impurity solver is constructed using it, as detailed in Sec.~\ref{subsec:varspaceEH}.

\subsubsection*{PCA compression as a foundation model}

The construction above shows that applying PCA compression to a data set consisting of ground states $\{\Phi(\mathbf{X}^s)\}$, sampled according to any probability distribution $p(\mathbf{X})$ over Hamiltonian parameters, is mathematically equivalent to determining an optimal variational subspace. Specifically, for a given $p(\mathbf{X})$, the resulting principal components span a shared variational subspace $\mathcal{V}_K$ that is optimal, in the sense of average fidelity, for all ground states generated by the family of embedding Hamiltonians $\hat{H}(\mathbf{X})$. 
In this sense, our PCA procedure provides a \emph{linear foundation model} for this family of Hamiltonians, in direct connection with variational foundation models based on neural-network quantum states that are trained to approximate ground states across continuous families of Hamiltonians~\cite{Rende-foundation-NNQS,Zhou2025NQS-impurity,Ma-DMET-NNQS}. The main difference is that here the cost functional is fidelity-based rather than energy-based. 

Our strategy described in Sec.~\ref{subsec:varspaceEH}, which consists of projecting the entire EH landscape $\{\hat{H}(\mathbf{X})\}$ into the learned variational space $\mathcal{V}_K$, is directly enabled by this conceptual connection between compression and foundation-model perspectives.
The resulting improvements in the efficiency of the overall QE procedure are discussed in Secs.~\ref{Sec:Benchmarks} and \ref{Sec:scalability}.

\section{Benchmark Calculations}
\label{Sec:Benchmarks}

In this section, we benchmark our PCA-based framework as follows. First in Sec.~\ref{ssec:3bandHubbmodel}, we benchmark it on the three-orbital Hubbard-Kanamori model using a pre-trained library of local variational spaces described in Sec.~\ref{ssec:PreTrainingApproach}, quantifying its accuracy and the resulting computational speedup. 
Second in Sec.~\ref{ssec:ActiveLearning}, we introduce an active-learning scheme that builds this library only on demand along the ghost-GA self-consistency procedure, showing that this substantially simplifies the learning process. Finally in Sec.~\ref{ssec:benchmarks-Pu}, we apply the same strategy to real materials, focusing on all allotropes of Pu under pressure, and show that the learned variational space captures the local $5f$ electronic environment and can be reused across all crystallographic phases.
% \og{Given this introduction, it makes more sense to me to have V A, B and C together rather than different subsections. B is giving details of A, and the title for C is literally the title of V + the title of A.}

\subsection{Degenerate 3-bands Hubbard-Kanamori models} \label{ssec:3bandHubbmodel}

\subsubsection{Lattice models}

To illustrate our approach, we first focus on solving model systems with 3 degenerate orbitals and with a local electron-electron Kanamori interaction~\cite{Kanamori1963,Georges2015_StrongCorrHund} and only nearest-neighbors hoppings. 
The Hamiltonian will generically read:
\begin{align}
    \hat{H} =& \hat{H}^0+\sum_{\mathbf{R}} \hat{H}^{int}_\mathbf{R}(U,J) \, ,\label{eq:lattice_hamiltonian}\\
    \hat{H}^0 =& -t\sum_{\langle \mathbf{R} \mathbf{R}^\prime\rangle } \sum_{\alpha \sigma} c^\dagger_{\mathbf{R}\alpha \sigma}c_{\mathbf{R}^\prime \alpha \sigma}-\mu \sum_{\mathbf{R} \alpha \sigma} c^\dagger_{\mathbf{R}\alpha \sigma} c_{\mathbf{R}\alpha \sigma} \,,
\end{align}
where $c^\dagger_{\mathbf{R} \alpha \sigma }$ ($c_{\mathbf{R} \alpha\sigma }$) is the electronic creation (annihilation) operator at site $\mathbf{R}$ for orbital $\alpha$ and spin $\sigma$, $\hat{H}^0$ is the non-interacting part of the Hamiltonian and $\hat{H}^{int}$ is the local Hubbard-Kanamori interaction.
We consider $U'=U-2J$, for which the interaction becomes rotationally invariant and will be parametrized by a Hubbard interaction $U$ and a Hund's coupling $J$. For each site $\mathbf{R}$, the interaction is defined as follows:
\begin{equation}
    \label{eq:int_ham_kanamori}
    \hat{H}^{int}_{\mathbf{R}}(U,J) = \frac{U-3J}{2}(\hat{N}_{\mathbf{R}})^2 -2J (\vec{S}_{\mathbf{R}})^2-\frac{J}{2} (\vec{L}_{\mathbf{R}})^2
    \,,
\end{equation}
where
\begin{align}
\hat{N}_{\mathbf{R}}&=\sum_{\alpha\sigma } c^\dagger_{\mathbf{R}\alpha \sigma} c_{\mathbf{R}\alpha \sigma}
\,,
\\
\vec{S}_{\mathbf{R}}&=\frac{1}{2}\sum_\alpha \sum_{\sigma \sigma^\prime}c^\dagger_{\mathbf{R}  \alpha \sigma}\vec{\tau}_{\sigma \sigma^\prime}c_{\mathbf{R}  \alpha \sigma} 
\,,
\\
{L}_{\mathbf{R},\gamma}&=i\sum_{\alpha \beta \sigma} \epsilon_{\gamma \alpha \beta}
\, c^\dagger_{\mathbf{R}  \alpha \sigma} c_{\mathbf{R}  \beta\sigma}
\,,
\end{align}
with $\vec{\tau}=(\tau_x,\tau_y,\tau_z)$ the vector of Pauli matrices and $\epsilon$ the Levi-Civita symbol.
The non-interacting part of the Hamiltonian in the basis for which the hopping term is diagonal:
\begin{equation}
    \hat{H}^0=\sum_{\alpha, \mathbf{k} \sigma} (\epsilon_\mathbf{k} -\mu) c^\dagger_{\mathbf{k} \alpha \sigma} c_{\mathbf{k} \alpha \sigma}
    \,,
\end{equation}
where $\epsilon_\mathbf{k}$ is the Fourier transform of the hopping Hamiltonian and $\mu$ is the chemical potential.
We will consider the Bethe lattice, featuring a semi-circular density of states $\rho(\epsilon)=({2}/{\pi W^2})\sqrt{W^2-\epsilon^2}$, the square lattice, for which $\epsilon_\mathbf{k}=-(W/2)(\cos k_x + \cos k_y)$, and the cubic lattice, for which $\epsilon_\mathbf{k}=-(W/3)(\cos k_x + \cos k_y+ \cos k_z)$, where $W$ is the half-bandwidth and we will express most of our results in units of $W$.
Once interactions are taken into account, the lattice model retains a charge-$U(1) $, spin-$SU(2)$ and orbital-$SO(3)$ symmetry.

\begin{figure}[h!]
    \centering
    \includegraphics[width=\linewidth]{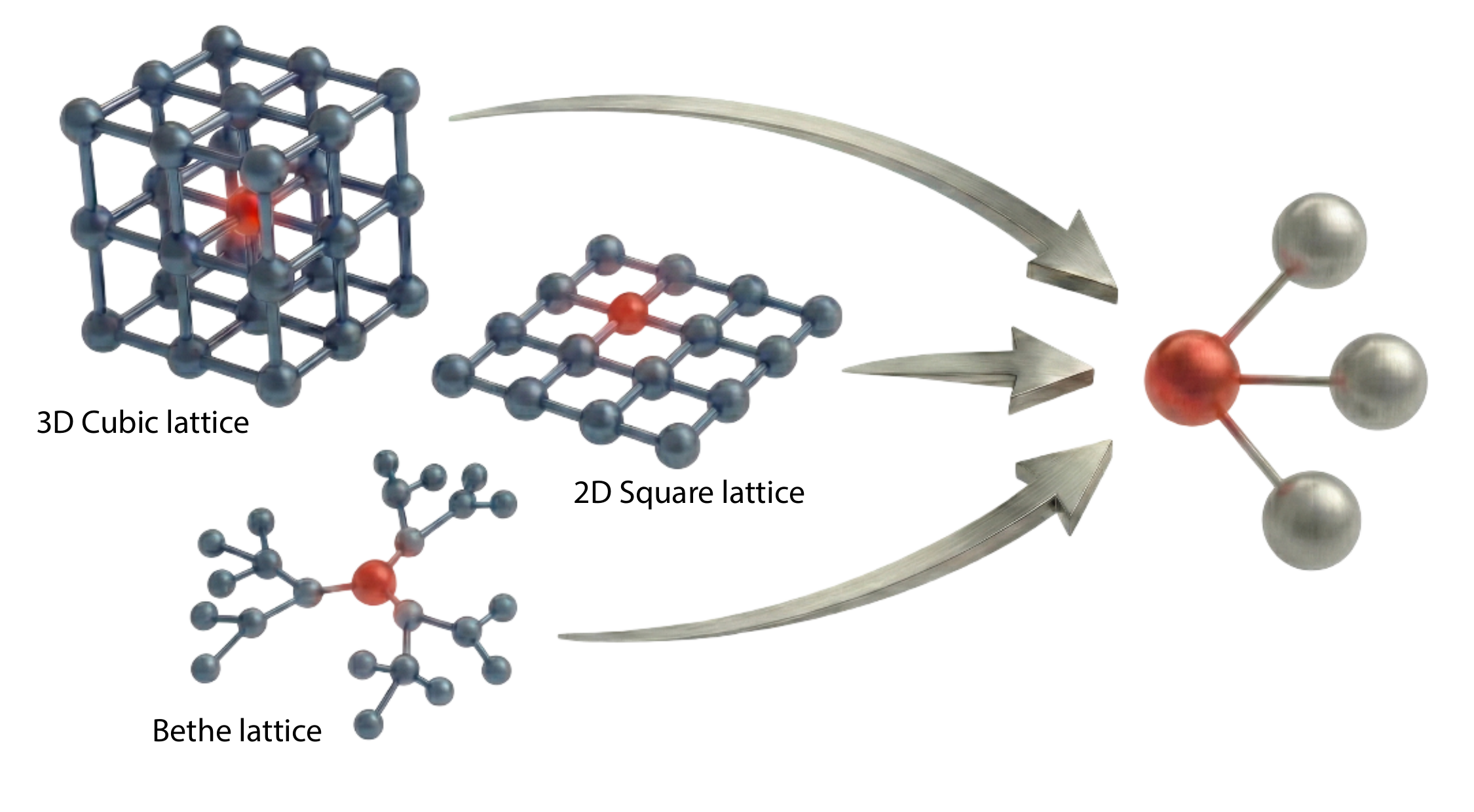}
    \caption{Pictorial representation of how different lattice models (cubic, square and Bethe) map onto an embedding Hamiltonian with the same structure.}
    \label{figure2}
\end{figure}

\subsubsection{Structure of the EH in the case of three degenerate orbitals}

The global symmetries of the lattice Hamiltonian will be reflected as symmetries of the EH~\cite{Ghost-GA,Lanata2016}. Different lattices with the same point-group symmetry will share the structure of the embedding Hamiltonian as shown in Fig.~\ref{figure2} and imposing those symmetries will drastically reduce the number of free parameters in the Hamiltonian. 
In particular, we impose that the ghost-GA variational wavefunction $|\Psi_G\rangle$ respects charge $U(1)$, spin $SU(2)$, and orbital $SO(3)$ symmetries. 
These constraints translate into corresponding restrictions on the EH~\cite{Our-PRX,Lanata2016}. The $U(1)$ constraint implies that the EH is at half filling. The $SU(2)$ constraint implies that the EH can be chosen spin-rotation invariant, with a spin-singlet ground state and a one-body part that splits into two equivalent spin-up and spin-down terms. 
Finally, the $SO(3)$ constraint implies that the one-body part of the EH contains no terms coupling different orbitals, and that all one-body parameters are identical for the three orbitals.
The Hamiltonian then reads:
\begin{align}
\hat{H} &= \hat{H}^{\text{int}}(U,J)
+ 
E \sum_{\alpha=1}^{3}\sum_{\sigma\in{\uparrow,\downarrow}} c_{\alpha \sigma}^\dagger c_{\alpha \sigma}
\nonumber\\
&+ \sum_{a=1}^{3}  D_a \sum_{\alpha=1}^{3} \sum_{\sigma\in{\uparrow,\downarrow}} (c_{\alpha \sigma}^\dagger b_{\alpha a \sigma} + \text{H.c.})  \nonumber \\
&  + \sum_{a,a^\prime=1}^{3}  
\Lambda_{a a^\prime} \sum_{\alpha =1}^{3} \sum_{\sigma\in{\uparrow,\downarrow}} b_{\alpha a \sigma}^\dagger b_{\alpha a^\prime\sigma} 
\,. \label{eq:3band_hamiltonian_postsym}
\end{align}
where $c_{\alpha \sigma}^\dagger$ ($c_{\alpha \sigma}$) creates (annihilates) an electron in the impurity $p$-orbital $\alpha \in \{1, 2, 3\}$ with spin $\sigma$, and $b_{\alpha a \sigma}^\dagger$ ($b_{\alpha a\sigma}$) creates (annihilates) an electron on the corresponding bath site in the bath-channel $a \in \{1, 2, 3\}$. The term $\hat{H}^{\text{int}}(U,J)$ is the standard two-body Kanamori interaction, whose matrix elements are determined by the parameters $U$ and $J$, as defined in Eq.~\ref{eq:int_ham_kanamori}. The parameters $D_a$ control the hybridization strength between the impurity and bath channel $a$, while $\Lambda_{aa^\prime}$ sets the on-site energies and hybridizations for all bath sites within that physical orbital channel.

To reduce the parameter space that must be covered by the ML model, we proceed as follows:
\begin{itemize}

\item Unitary canonical transformations acting among the bath orbitals do not affect the hybridization function and, in turn, the physical interaction between the impurity and the bath. Within the ghost-GA formalism, such transformations are referred to as gauge transformations. 
Following Ref.~\cite{surrogate-Marius}, we use gauge transformations to restrict our ML task to a bath representation with $\Lambda$ diagonal. Furthermore, we sort the eigenvalues of $\Lambda$ in ascending order ($\Lambda_a<\Lambda_{a+1}$), and choose the hybridization parameters $D_a$ to be real and negative.

\item Multiplying the EH by an overall positive factor does not change its ground-state eigenvector. 
Therefore, without loss of generality we fix the energy unit by setting $U=1$ in the EH parametrization used for the ML task.

\item As explained above, the charge $U(1)$ symmetry implies that the EH ground state has to be calculated within the half-filled sector. 
Since adding to the EH a chemical-potential term proportional to the total number of fermions (including bath and impurity) only shifts all eigenvalues by a constant, but does not change eigenvectors, we can set $E=0$ with no loss of generality.

\end{itemize}
With these simplifications, the EH reads:
\begin{align}
\hat{H} &= \hat{H}^{\text{int}}(U,J) + 
\sum_{a=1}^{3} \Lambda_a \sum_{\alpha=1}^{3} \sum_{\sigma\in{\uparrow,\downarrow}} b_{\alpha a \sigma}^\dagger b_{\alpha a \sigma} \nonumber \\
&+ \sum_{a=1}^{3}  D_a \sum_{\alpha=1}^{3} \sum_{\sigma\in{\uparrow,\downarrow}} (c_{\alpha\sigma}^\dagger b_{\alpha a\sigma} + \text{H.c.}) 
\,,
\label{eq:3band_hamiltonian}
\end{align}
which can be rewritten as follows:
\begin{equation}
\hat{H}(\mathbf{X}) = \hat{H}^{\text{int}}(U,J) + \sum_{r=1}^{6} X_r \hat{H}_r
\,, \label{eq:Hemb_3deg}
\end{equation}
where $\hat{H}_r$ denote the hybridization and bath-energy terms in Eq.~\eqref{eq:3band_hamiltonian}, and the parameter vector is
\begin{equation}
\mathbf{X} = (D_1, D_2, D_3, \Lambda_1, \Lambda_2, \Lambda_3)\,. \label{eq:X_deg}
\end{equation}
As discussed above, we set $U=1$. 

Since $J$ is not updated throughout the gGA self-consistency procedure, from now on we set $J/U=0.1$ for simplicity.

\subsection{Pre-training approach}
\label{ssec:PreTrainingApproach}

In this section, we construct an atlas of local charts in the parameter space defined in Eq.~\eqref{eq:X_deg}. The atlas is built over a prescribed parameter domain of embedding-Hamiltonian parameters \(\mathbf{X}\). Each chart is assumed to be hyperrectangular, characterized by a center \(\mathbf{X}_j\) and by fixed linear sizes \(L_D\) and \(L_\Lambda\) along the \(D\) and \(\Lambda\) directions, respectively.

To build the atlas, we repeatedly draw random samples \(\mathbf{X}_j\) from a uniform distribution over the target domain. If a sampled point \(\mathbf{X}_j\) lies inside at least one chart that has already been learned, we discard the sample. Otherwise, we define a new chart centered at \(\mathbf{X}_j\) (with the sizes specified above) and associate to it a variational solver. We iterate this procedure until additional random samples no longer generate new charts, at which point the atlas provides a covering of the chosen domain.

\subsubsection{Training procedure for a single chart}
\label{sssec:charts_learning_gga}

Throughout this section we fix \(L_D=0.4\) and \(L_\Lambda=4.0\).
For each chart, we follow the PCA construction described in Sec.~\ref{Sec:learning} using a finite local training set. Specifically, we draw \(N_{\mathrm{budget}}\) parameter vectors \(\{\mathbf{X}^s\}\) uniformly within the chart, compute the corresponding EH ground states, and perform the SVD of the resulting data matrix. In our calculations we set \(N_{\mathrm{budget}}=600\).

We then determine dynamically the dimension \(K\) of the local variational space from a prescribed accuracy threshold, which we set to \(\varepsilon_{\mathrm{tol}}=10^{-6}\). 
Concretely, we set
\begin{equation}
K=\min\left\{k \,\big|\, \sigma_{k}/\sigma_1 < \varepsilon_{\mathrm{tol}}\right\}
\,,
\label{eq:cutoff_condition}
\end{equation}
where \(\{\sigma_k\}\) are the singular values sorted in descending order. This criterion is motivated by the fact that the singular values quantify the weight of each principal direction in representing the training set: if \(\sigma_k\) decays rapidly with \(k\), the sampled ground states lie close to a low-dimensional linear subspace. Requiring \(\sigma_{K}/\sigma_1 < \varepsilon_{\mathrm{tol}}\) therefore ensures that principal components beyond \(K\) carry a negligible weight compared to the leading one, so that the span of the first \(K\) components provides an accurate compressed representation of the training data within the chart.

Since the data matrix $\mathbb{M}$ has at most rank \(N_{\mathrm{budget}}\), the SVD spectrum is resolved only up to \(k=N_{\mathrm{budget}}\). For this reason, if the value of \(K\) determined by Eq.~\eqref{eq:cutoff_condition} approaches \(N_{\mathrm{budget}}\), it becomes sensitive to the finite sampling budget and cannot be interpreted as a robust estimate of the intrinsic truncation dimension. To guard against this undersampling effect, we impose the additional buffer condition \(K/N_{\mathrm{budget}} < 0.8\). If this condition is not satisfied, we keep \(N_{\mathrm{budget}}\) fixed and replace the chart by a smaller one with the same center \(\mathbf{X}_j\) but with \(L_D\) and \(L_\Lambda\) reduced by a factor 2, and repeat the sampling and PCA procedure. This process is iterated until the singular-value decay is sufficiently well resolved at the target tolerance.

Once the condition \(K/N_{\mathrm{budget}} < 0.8\) is met, we retain the first \(K\) principal components \(\{|u_n\rangle\}_{n=1}^{K}\), which form an orthonormal basis of the compressed variational space used to represent EH ground states within the chart. 
We then compute and store the corresponding matrix elements in this basis for the operators needed within the ghost-GA loop.
In the present symmetry-reduced setting, these can be taken as
\begin{align}
[\tilde{O}^h_{\alpha a}]_{nm} &= 
\sum_{\sigma}\langle u_n|c_{\alpha\sigma}^\dagger b_{\alpha a\sigma}|u_m\rangle
\,,
\label{eq:Oh}
\\
[\tilde{O}^d_{\alpha a}]_{nm} &= 
\sum_{\sigma}\langle u_n|b_{\alpha a\sigma}^\dagger b_{\alpha a\sigma}|u_m\rangle
\,,
\label{eq:Od}
\\
[\tilde{H}^{\text{int}}]_{nm} &= 
\langle u_n|\hat{H}^{\text{int}}|u_m\rangle
\,.
\label{eq:Oi}
\end{align}
This precomputation allows subsequent EH solutions with \(\mathbf{X}\) lying within the trained chart to be carried out as a deterministic \(K\times K\) eigenvalue problem within the corresponding \(K\)-dimensional variational space.

%%%%%%%

\subsubsection{Pre-training construction and atlas statistics}
\label{sssec:PreTraining_atlas}

For the degenerate three-orbital problem introduced in Sec.~\ref{ssec:3bandHubbmodel}, we take the pre-training domain to be \(D_i \in [-1.0,0.0]\) and \(\Lambda_i \in [-10.0,10.0]\) for \(i=1,2,3\).

Applying the procedure above across this pre-training domain yields an atlas consisting of 990 charts. In practice, we find that the atlas construction saturates after \(\sim 10^4\) random trial centers, in the sense that additional samples no longer generate new charts. 

\begin{figure}
    \centering
    \includegraphics[width=1.0\linewidth]{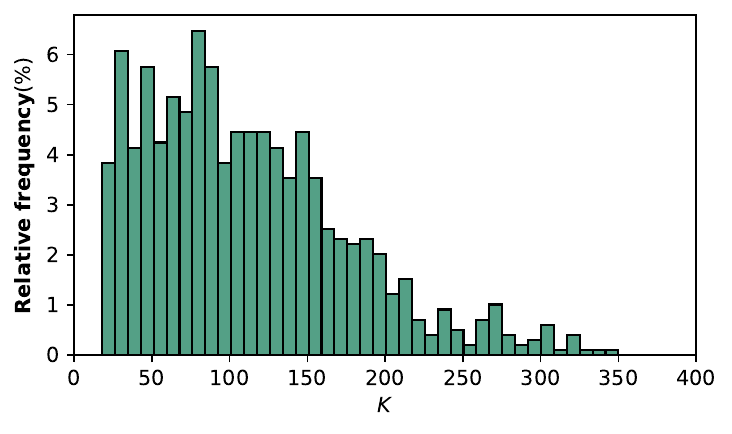}
    \caption{Distribution of the number of principal components retained to define the local variational space across the 990 charts of the pre-trained atlas described in Sec.~\ref{sssec:PreTraining_atlas}.}
    \label{figure3}
\end{figure}

Across all charts, the resulting distribution of variational dimensions \(K\) is shown in Fig.~\ref{figure3}; the largest value selected by the criteria above is \(\max(K)=350\), compared to a Fock-space dimension of \(853{,}776\) for the full EH. The resulting database of projected operators for all charts has size \(1.75\)~GB.

\begin{figure*}
    \centering
    \includegraphics[trim={5cm 0 5cm 5cm},clip,width=1.0\linewidth]{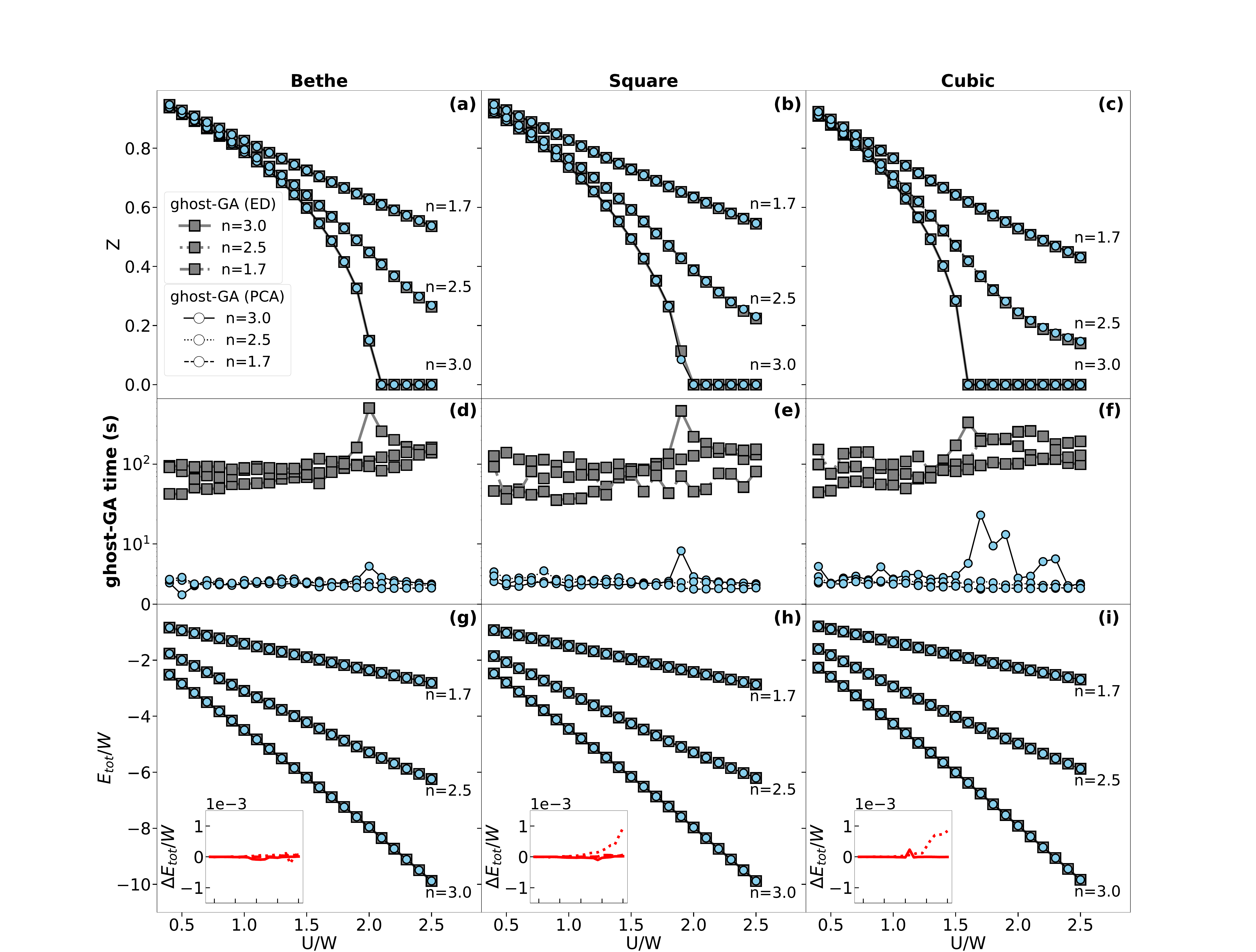}
    \caption{Ghost-GA results for the three degenerate orbitals Hubbard-Kanamori model with $J/U=0.1$ on the Bethe (square, cubic) lattice for the left (center, right) column using pre-trained PCA (circles) and ED (squares) as EH solvers. (a,b,c) Quasiparticle weight, (d,e,f) ghost-GA calculation time and (g,h,i) total energy per site as a function of $U/W$ at filling $n=3.0, 2.5,1.7$. The inset show the difference between the total ghost-GA energies per site computed with ED and PCA solvers for the EH.
    }
    \label{figure4}
\end{figure*}

\subsubsection{Ghost-GA results with the pre-trained solver}

Having constructed the atlas and the corresponding database of projected operators, we next use the resulting compressed solver within full ghost-GA calculations for the three-orbital Hubbard--Kanamori model introduced in Sec.~\ref{ssec:3bandHubbmodel}. We benchmark ghost-GA calculations in which the EH is solved using the pre-trained PCA solver against reference calculations in which the EH is solved by exact diagonalization (ED).

During the ghost-GA self-consistency cycle, the EH parameters \(\mathbf{X}\) at a given iteration may fall within multiple charts of the pre-trained atlas. In such cases, we exploit the variational character of the projected solver: we evaluate the EH ground-state energy in each admissible chart and select the solution yielding the lowest energy.

Figure~\ref{figure4} compares ghost-GA results obtained with the ED solver (squares) and with the pre-trained PCA solver (circles) for the degenerate three-orbital model as a function of \(U/W\) on the Bethe, square, and cubic lattices (left, center, and right columns, respectively), and for the fillings indicated in the figure. The quasiparticle weight \(Z\), Figs.~\ref{figure4}(a--c), and the total energy per site, Figs.~\ref{figure4}(g--i), are reproduced with high accuracy by the PCA solver. The insets of Figs.~\ref{figure4}(g--i) report the corresponding energy differences \(E_{\mathrm{ED}}-E_{\mathrm{PCA}}\), which remain small across the explored interaction range. At the same time, the total wall time of the ghost-GA cycle is reduced substantially when using the PCA solver, as shown in Figs.~\ref{figure4}(d--f), with speedups ranging from one order of magnitude to nearly two orders of magnitude in the vicinity of the Mott transition.

These benchmarks demonstrate that a solver pre-trained once for a given EH structure can be reused across different lattice realizations of the same local problem, delivering both robust accuracy and significant computational acceleration. This reduction of the EH cost directly enhances the feasibility of systematic scans and high-throughput studies within ghost-GA.

\subsection{Active Learning Approach}
\label{ssec:ActiveLearning}

In the previous subsection, we showed that constructing an atlas of local variational solvers can substantially accelerate quantum-embedding calculations by reducing the cost of the EH step. However, as shown in Ref.~\cite{surrogate-Marius}, the physically relevant EHs populate a sparse, low-dimensional manifold, rather than being spread throughout the full parameter domain. Consequently, globally pre-training a solver over a large parameter region can waste significant computational effort on regions that are never visited in QE calculations.

Motivated by this observation, we adopt an active-learning (on-demand) strategy in which the database is initialized as empty and updated only when needed. During the self-consistency cycle, if the current EH parameter \(\mathbf{X}\) is contained in at least one previously learned chart, we solve the EH using the corresponding PCA-projected solver; otherwise, we solve the EH by exact diagonalization (ED) and record the uncovered \(\mathbf{X}\) values. After convergence at a given set of physical parameters, any newly recorded \(\mathbf{X}\) values are used as training centers to construct additional charts following Sec.~\ref{sssec:charts_learning_gga}, and the resulting projected operators are appended to the database before proceeding.

\begin{figure*}
    \centering
    \includegraphics[trim={7cm 0cm 10cm 5cm},clip,width=\textwidth]{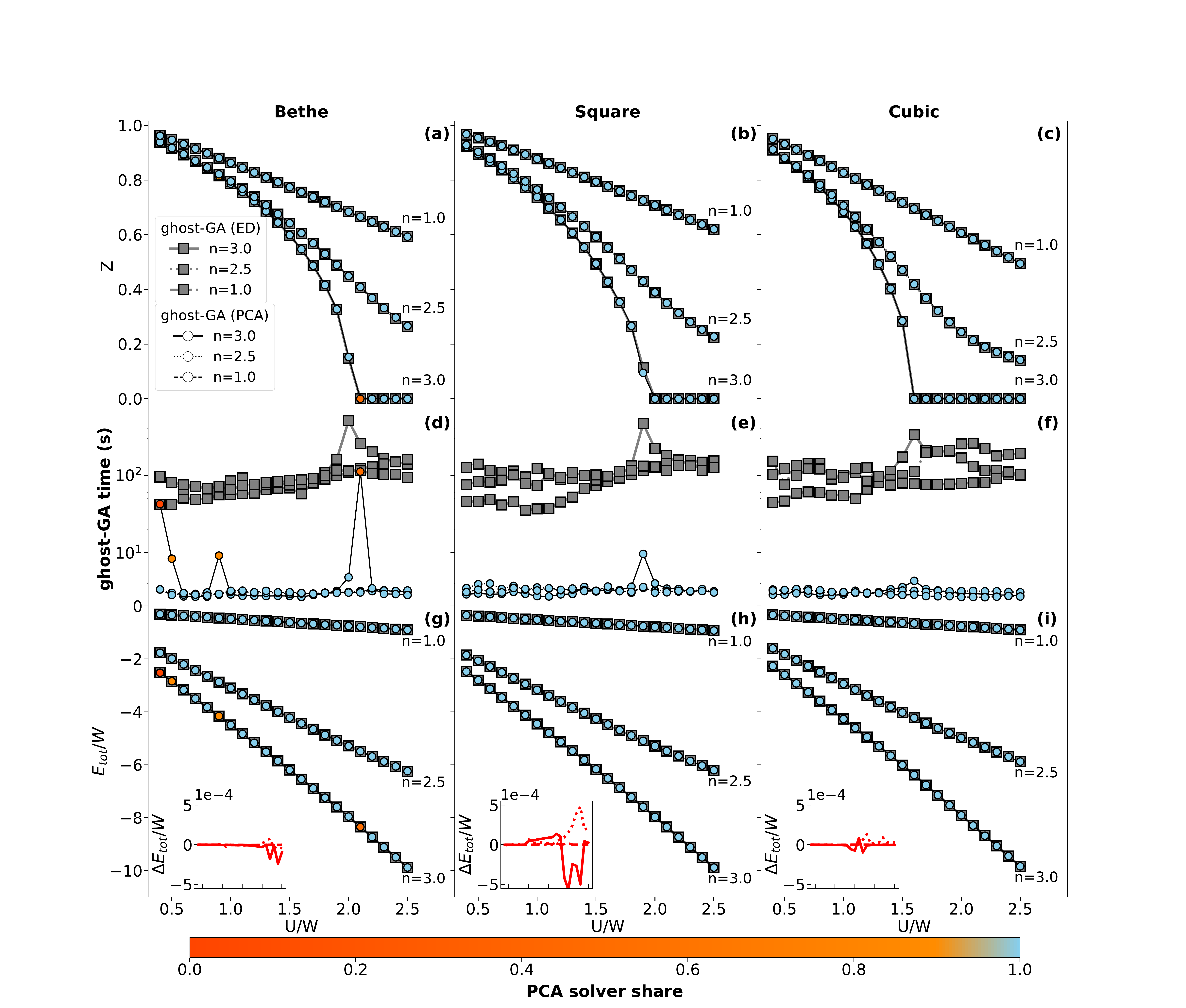}
    \caption{Ghost-GA results for the three degenerate-orbital Hubbard--Kanamori model with \(J/U=0.1\) on the Bethe (square, cubic) lattice in the left (center, right) column using the active-learning PCA solver (circles) and ED (squares) as EH solvers. (a--c) Quasiparticle weight, (d--f) ghost-GA calculation time, and (g--i) total energy per site as a function of \(U/W\) at fillings \(n=3.0,2.5,1.0\). Insets: differences between the total energies per site computed with ED and PCA solvers for the EH. Marker color indicates the fraction of EH solves within the ghost-GA cycle handled by the PCA solver (red: 0; blue: 1).}
    \label{figure5}
\end{figure*}

We benchmark this approach by running ghost-GA calculations over the combined set of choices of lattice realization, electron filling \(n\), and interaction strength \(U\). We first perform the calculations on the Bethe lattice (left column of Fig.~\ref{figure5}), starting from an empty database, and then apply the database accumulated on the Bethe lattice to the square and cubic lattices (center and right columns). In all cases, we traverse the \((n,U)\) grid as follows: we loop over \(n=3.0,2.9,\ldots,0.5\). For each value of \(n\), we then sweep \(U/W=0.4,0.5,\ldots,2.5\).

On the Bethe lattice at half filling (\(n=3.0\)), Fig.~\ref{figure5}(a,d,g) compares ghost-GA calculations using the ED solver for the EH (squares) and the active-learning PCA solver (circles). At \(U/W=0.4\), the database is empty, so EH calls are solved by ED (red markers), which triggers chart creation. Already at \(U/W=0.5\), the learned charts cover almost all EH calls (blue markers). Over the half-filled sweep, the EH solutions are learned with four charts created at \(U/W=0.4,0.5,0.9,2.1\). The corresponding deviations in the total energy are shown in the inset of Fig.~\ref{figure5}(g) and are four orders of magnitude smaller than the bandwidth \(W\).

Continuing the Bethe-lattice calculations down to \(n=0.5\), in the first column of Fig.~\ref{figure5} we report representative results for \(Z\) [Fig.~\ref{figure5}(a)], ghost-GA calculation time [Fig.~\ref{figure5}(d)], and total energy [Fig.~\ref{figure5}(g)] at \(n=3.0,2.5,1.0\). After the initial learning at half filling, only a small number of additional charts is required; in the present calculations the atlas grows to a total of 13 charts. The resulting database of compressed solvers has size \(\sim 100\)~MB, corresponding to an order-of-magnitude reduction in memory footprint with respect to the pre-training approach.

Using the database accumulated on the Bethe lattice, the center (square) and right (cubic) columns of Fig.~\ref{figure5} report the corresponding results for lattices with different geometries. For the data points considered here, no additional training steps are required. In both cases, the error in the total energy remains four orders of magnitude smaller than the half-bandwidth, while the overall ghost-GA calculation time is reduced by approximately a factor \(30\) [Fig.~\ref{figure5}(e,f)].

Overall, active learning reduces the number of required charts from 990 (global pre-training) to 13 in this scan, yielding substantial savings in both computational cost and memory footprint, while retaining accuracy and transferability.

\subsection{Active-learning application to all phases of Pu}
\label{ssec:benchmarks-Pu}

In this section, our aim is to assess the performance benefits afforded by the PCA-based active-learning framework in a realistic materials context, where the environments of different structural fragments—here encoded in the embedding Hamiltonians (EHs)—are constrained by physical factors such as crystal symmetry, structural stability, and the nature of chemical bonding. To this end, we apply the PCA-compressed ghost-GA solver to a concrete real-materials problem: computing the energy–volume $E(V)$ curves of six crystalline phases of Pu with a \textit{single} trained solver reused for all volumes and phases.

In the Pu systems, the f-electron states are split by spin–orbit coupling (SOC) and crystal-field splitting (CFS). Since SOC is dominant, the standard CFS-averaged approximation is adopted, where the f shell is treated as isotropic within the EH, and correlation effects distinguish only the $J=5/2$ and $J=7/2$ manifolds. We set $B=1$, $U=4$~eV and $J/U=0.09$ for the practical calculations of all the Pu phases~\cite{Our-PRX}. The embedding Hamiltonian takes the following form:
\begin{widetext}
\begin{equation}
\begin{aligned}
\hat{H}_{\text{emb}}(\mathbf{X})=\hat{H}_{\text{int}}(U,J)
&+ E_{5/2}\sum_{j_z=-5/2}^{5/2} c_{5/2,j_z}^\dagger c_{5/2,j_z}
+ E_{7/2}\sum_{j_z=-7/2}^{7/2} c_{7/2,j_z}^\dagger c_{7/2,j_z}
\\
&+ \epsilon_{5/2}\sum_{j_z=-5/2}^{5/2} b_{5/2,j_z}^\dagger b_{5/2,j_z}
+ \epsilon_{7/2}\sum_{j_z=-7/2}^{7/2} b_{7/2,j_z}^\dagger b_{7/2,j_z}
\\
&+ v_{5/2}\sum_{j_z=-5/2}^{5/2}\!\!\left(c_{5/2,j_z}^\dagger b_{5/2,j_z}+\textrm{H.c.}\right)
+ v_{7/2}\sum_{j_z=-7/2}^{7/2}\!\!\left(c_{7/2,j_z}^\dagger b_{7/2,j_z}+\textrm{H.c.}\right),
\end{aligned}
\end{equation}
\end{widetext}
with independent parameters $\mathbf{X}=(E_{5/2}-E_{7/2}, \,v_{5/2}, \allowbreak  \,v_{7/2},  \,\epsilon_{5/2},\,\epsilon_{7/2})$.

The EH structure above is identical for every Pu phase and volume considered. The training and inference therefore always occur in the same symmetry sector with fixed half filling at $N_\textrm{e} = 14$ and total angular momentum $J=0$ for the impurity-bath system. In numerical ED calculations, we employ a Fock-state basis with $N_\textrm{e} = 14$ and total angular momentum projection $J_\textrm{z} = 0$, resulting in a basis set dimension of $2{,}831{,}584$. 

\begin{figure}
    \centering
    \includegraphics[width=0.48\textwidth]{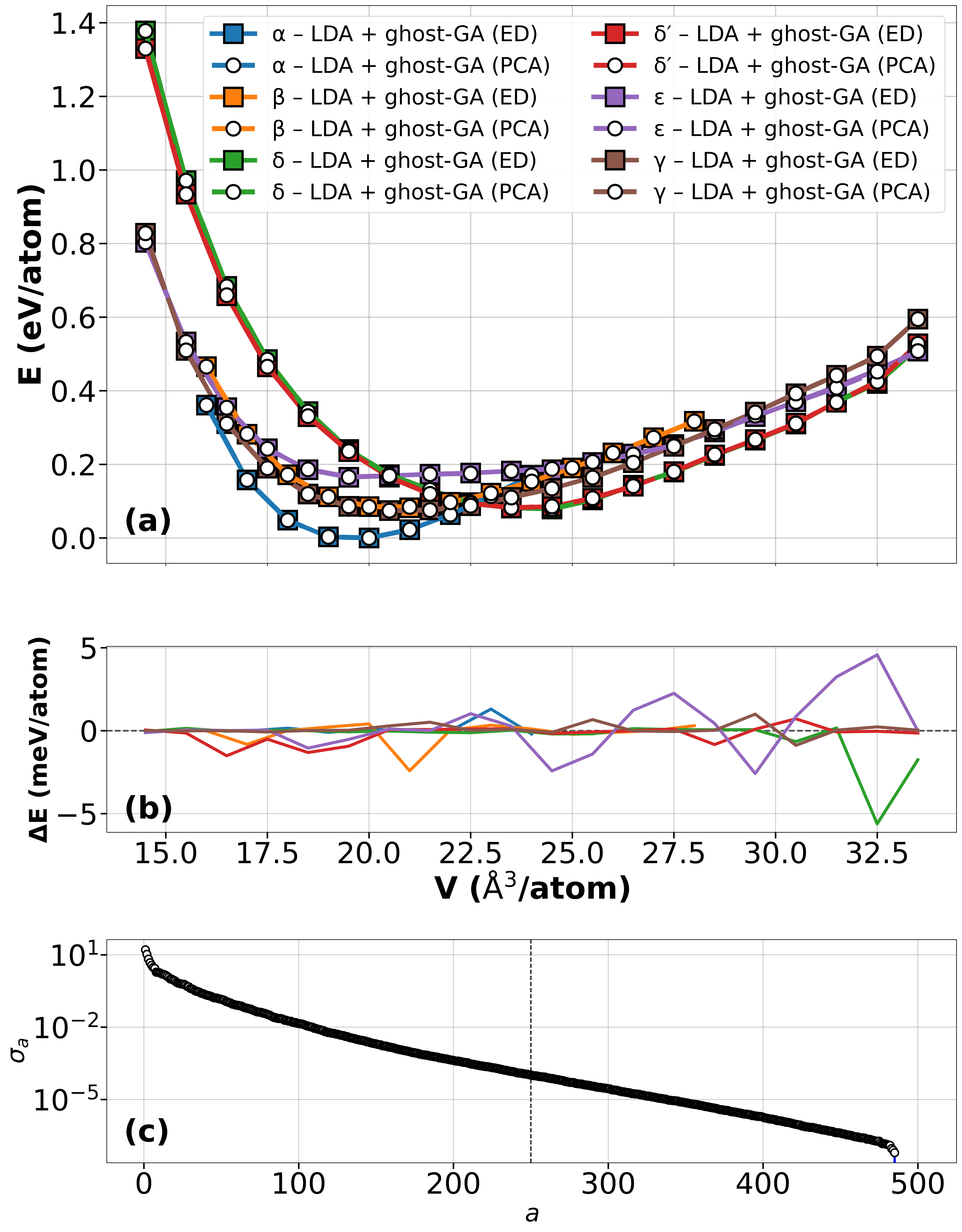}
    \caption{(a) Energy–volume curves for the $\alpha$, $\beta$, $\delta$, $\delta'$, $\varepsilon$, and $\gamma$ phases of Pu. Filled markers: LDA+ghost-GA with exact ED solver. Open markers: LDA+ghost-GA with the same PCA-compressed solver trained once on a fixed EH-parameter box and reused for all phases and volumes. (b) The energy difference between the calculations with ED and PCA solvers, $\Delta E = E_\textrm{ED} - E_\textrm{PCA}$, for each of the six phases.
    (c) Singular values $\sigma_a$ of matrix $\mathbb{M}$ in descending order.
    }
    \label{figure6}
\end{figure}

We construct the training set within a hyper-rectangular box in the 5D parameter space of $\mathbf{X}$. The range of $\mathbf{X}$ is chosen to be about twice the parameter value range in self-consistent LDA+ghost-GA calculation for $\delta$-Pu at a minimal volume ($14.5\text{ \AA}^3/\text{atom}$) and maximal volume point ($33.5\text{ \AA}^3/\text{atom}$). Specifically, the range is $[0.25, 0.29]$ for $(E_{5/2}-E_{7/2})/U$, $[-1, 0]$ for $v_{5/2}/U$ and $v_{7/2}/U$, $[3.38, 6.10]$ for $\epsilon_{5/2}/U$, and $[2.92, 5.48]$ for $\epsilon_{7/2}/U$. From this hyper-rectangular box we randomly draw $N=500$ sets of parameters to construct EH instances, compute their exact ground states via ED, and apply PCA analysis to the resulting column-wise stacked state vectors. We retain $K=250$ principal components to define the variational subspace with the singular value $\sigma_a \ge 10^{-4}$, as shown in Fig.~\ref{figure6}(c). The projected operators $\{\tilde H_r^{(K)}\}$ are precomputed once and cached. The variational subspace is fixed and reused for every EH solution in the LDA+ghost-GA calculations of all volumes and across all Pu phases without retraining, patching, or basis updates.

In Fig.~\ref{figure6}(a) we compare LDA+ghost-GA $E(V)$ curves obtained with the ED solver and with the single PCA-compressed solver for the $\alpha$, $\beta$ $\delta$, $\delta'$, $\varepsilon$, and $\gamma$ phases. Across the entire sampled volume range the two solvers are indistinguishable at the plot scale. The volume-wise energy differences are shown in In Fig.~\ref{figure6}(b), which are mostly smaller than $5$\,meV per atom.

Because the same $K=250$ basis is used for every EH call, the savings are uniform across phases and volumes. Replacing ED with the PCA solver reduces the cumulative time spent inside the EH routine by a factor $\sim 4000$ ($\sim 200$ min vs $0.05$ min), while substantially lowering memory usage by diagonalizing a Hamiltonian in the $K\times K$ subspace instead of the full Hilbert space. Thus, a single training on one 5D box yields ED-level accuracy for all considered Pu phases with orders-of-magnitude reductions in CPU time and RAM.

\section{Scalability and Integration with High-Level Solvers}
\label{Sec:scalability}

The results presented thus far establish that the ground-state manifold of the embedding Hamiltonian admits a highly compact representation. This property has enabled us to drastically accelerate simulations of f-electron materials and multi-orbital models using training data generated by exact diagonalization. However, extending this methodology to regimes governed by even larger Hilbert spaces requires overcoming the exponential scaling of the basis size. This challenge arises, for example, in ghost-GA calculations with larger bath sizes ($B \ge 3$), such as for transition metal oxides~\cite{TH3} or cluster impurities, where the dimension of the embedding Hamiltonian precludes the use of Arnoldi solvers. In these regimes, one must rely on high-level approaches like MPS~\cite{itensor,block2,DMRG-REVIEW,DMRG-original-White-PRL,DMRG-original-White-PRB,DMRG_PhysRevB.104.115119}, NQS, or variational impurity solvers based on superpositions of Gaussian states~\cite{Bauer-impurity-solver,Wu2025_gaussian}.

While these solvers can effectively handle larger systems, integrating them with the PCA framework described in Sec.~IV poses a specific technical hurdle: standard PCA requires the explicit storage of full wavefunction vectors, which is often infeasible or computationally prohibitive for high-level representations. To unlock the capability of our framework to interface with these advanced solvers, it is necessary to implement a generalized formulation that remains agnostic to the underlying data structure. In this section, we introduce the ``method of snapshots'', which constructs the variational space relying solely on state overlaps, providing a general pathway for scalable quantum embedding simulations.

\subsection{Training PCA models from general impurity solvers using the ``method of snapshots''}
\label{Sec:method-snapshots}

The core observation underlying this generalization is that the construction of the PCA variational space does not fundamentally require explicit state vectors, but only access to the overlaps between training states. This naturally leads to the formulation known as the ``method of snapshots,'' originally introduced in the context of reduced-order modeling in fluid dynamics, which recasts the diagonalization problem in terms of the overlap matrix in the space of samples rather than in the full Hilbert space. By lifting the restriction to solvers that explicitly return the wavefunction as a column vector, this reformulation makes the PCA-based solver compatible with any impurity solver capable of computing overlaps between ground states.

First, we diagonalize the overlap matrix of the sampled states:
\begin{equation}
\sum_{\alpha,\beta=1}^{N}V_{a\alpha}^\dagger\langle\Phi(\mathbf{X}^\alpha)|\Phi(\mathbf{X}^\beta)\rangle V_{\beta b} = \sigma_a^2\delta_{ab}
\,,
\label{eq:snapshots-part1}
\end{equation}
where $a,b = 1,\ldots,N$, and $V$ is the matrix of eigenvectors. Note that this matrix is only $N\times N$, and so it is considerably smaller than the dimension of the Hilbert space $M_B$.
Second, this diagonalization produces a set of orthonormal vectors:
\begin{equation}
|u_a\rangle = \sigma_a^{-1}\sum_{\alpha=1}^{N}V_{\alpha a}|\Phi(\mathbf{X}^\alpha)\rangle
\,,
\label{eq:snapshots-part2}
\end{equation}
with $\sigma_a$ being the singular values sorted in descending order.
This approach provides an indirect but exact way of computing the singular values and principal components of the training data, bypassing the need to store wavefunctions as explicit $M_B$-dimensional arrays.

Significantly, Eqs.~\eqref{eq:snapshots-part1}--\eqref{eq:snapshots-part2} require only the overlaps $\langle\Phi(\mathbf{X}^\alpha)|\Phi(\mathbf{X}^\beta)\rangle$ and linear combinations of the corresponding states. As a result, the same PCA-based EH reduction can be trained on data generated by virtually any advanced impurity solver, including MPS-based methods or quantum-assisted algorithms~\cite{Sriluckshmy2025,Chen2025,Error-mitigation-GPR}. In this way, the PCA framework presented here becomes a general reduction layer that can sit on top of a wide class of solvers, amortizing their cost over many ghost-GA calculations.

The formulation of the PCA solver via the method of snapshots, Eqs.~\eqref{eq:snapshots-part1}--\eqref{eq:snapshots-part2}, essentially decouples the generation of the variational basis from the solution of the embedding problem. This separation offers a strategic advantage when integrating with emerging NQS solvers. Recent benchmarks of NQS impurity solvers for quantum embedding have identified that, while the variational optimization of the wavefunction is computationally efficient, the subsequent "inference" step ---sampling physical observables with the high precision required for self-consistency--- remains a primary computational bottleneck~\cite{Zhou2025NQS-impurity}.
Our PCA framework may provide a direct resolution to this inference problem. Instead of requiring the NQS to sample observables repeatedly during the self-consistency loop, one can use the NQS solely as a high-quality generator of basis snapshots. Once these states are projected into the compressed PCA subspace via the method of snapshots, all subsequent observable calculations become deterministic matrix operations within the low-dimensional subspace. This effectively shifts the heavy computational burden of Monte Carlo sampling entirely to the pre-training phase, leaving a fast, noise-free solver for the embedding loop.

\subsection{Integration with Neural-Network Quantum States}
\label{sec:future-integration-with-nns}

Looking beyond the generation of snapshots, the existence of a compact variational basis suggests the potential for more advanced hybrid architectures. Recent foundation models based on neural networks have demonstrated the capability to learn ground-state approximations across continuous families of Hamiltonians by taking the physical couplings as explicit inputs~\cite{Rende-foundation-NNQS,Zhou2025NQS-impurity,Ma-DMET-NNQS}. A promising future direction lies in combining these neural-network approaches with the linear subspace structure identified in this work. By explicitly exploiting problem-specific mathematical features, such as the affine dependence of the embedding Hamiltonian on its parameters, it may be possible to construct hybrid variational ansätze that learn the optimal linear basis and the mixing coefficients simultaneously. Such strategies, combining the representational power of neural networks with the physical robustness of linear projections, may enable simulations of strongly correlated systems with even larger and more complex embedding problems.

\section{Conclusions}

In this work, we introduced a linear foundation model for ghost-GA based on learning low-dimensional variational spaces for families of embedding Hamiltonians. Using PCA, we constructed local linear subspaces adapted to the EH ground-state manifold and projected the EH onto these subspaces. Once the projected operators are precomputed, each EH problem is reduced to a small deterministic eigenproblem, enforcing by construction $N$-representability conditions~\cite{N-Representability-Coleman} (e.g., the Pauli principle).
Furthermore, we developed an active learning strategy to generate the variational basis ``on demand'', avoiding redundant training.
We benchmarked the method on multi-orbital Hubbard models and on all crystalline phases of Pu.  
In the Hubbard model, we showed that a database trained on the Bethe lattice could be transferred to square and cubic lattices without additional training, requiring only a small number of local domains (thirteen) to cover the entire phase diagram. In the case of Plutonium, the efficiency was even more pronounced: a single training box proved sufficient to describe the ground states across all six allotropes. This reflects that our method effectively captures the intrinsic similarity of the local 5f-shell environment across different structural phases.
By extracting the universal features of the local chemical bonding from a single representative domain, the solver avoids redundant training and generalizes across the phase diagram, reproducing ghost-GA total energies obtained with full ED impurity solvers to within a few meV per atom, while reducing the cost of the EH step by orders of magnitude.
These applications leverage the fact that the physically relevant embedding Hamiltonians, representing the interactions between fragments and their environments, span a much more limited range within the full parameter space
---which is the same physical principle underlying scalability in foundation models for universal interatomic potentials~\cite{Allen2024_FMMLIP,Kim2025_SevenNetMF,Ju2025_UniversalMLIP}.
Finally, we showed that this PCA construction can be reformulated via the method of snapshots, enabling the integration of data from high-level solvers such as MPS~\cite{itensor,block2,DMRG-REVIEW,DMRG-original-White-PRL,DMRG-original-White-PRB,DMRG_PhysRevB.104.115119} and neural-network quantum states~\cite{Zhou2025NQS-impurity,Ma-DMET-NNQS,Rende-foundation-NNQS}.
Overall, our results establish that the essential information required for the embedding loop can be efficiently learned. By compressing the relevant Hilbert space, we can effectively eliminate the primary computational bottleneck of the QE frameworks, opening a path toward accurate high-throughput \textit{ab initio} investigations of strongly correlated matter.

\section*{Acknowledgments}

N.L. gratefully acknowledges funding from the National Science Foundation under Award No. DMR-2532771 and from the Simons Foundation (Grant No. 00010910). The Flatiron Institute is a division of the Simons Foundation.
Part of this work by Y.Y. was supported by the US Department of Energy (DOE), Office of Science, Basic Energy Sciences, Materials Science and Engineering Division, including the grant of computer time at the National Energy Research Scientific Computing Center (NERSC) in Berkeley, California. This part of research was performed at the Ames National Laboratory, which is operated for the US DOE by Iowa State University under Contract No. DE-AC02-07CH11358. T.-H.L. gratefully acknowledges funding from the National Science and Technology Council (NSTC) of Taiwan under Grant No. NSTC 112-2112-M-194-007-MY3 and the National Center for Theoretical Sciences (NCTS) in Taiwan.

% \bibliography{ref}


\end{document}